\documentclass[arps,prd,twocolumn,twoside,floatfix,noshowpacs,nofootinbib,hyperref]{revtex4-1}
\pdfoutput=1

\usepackage{xcolor}
\usepackage{amsmath}
\usepackage{ulem}

\usepackage{bm}

\newcommand\beq{\begin{equation}}
\newcommand\eeq{\end{equation}}
\newcommand\beqn{\begin{eqnarray}}
\newcommand\eeqn{\end{eqnarray}}

\newcommand{\ba}{\begin{eqnarray}}
\newcommand{\ea}{\end{eqnarray}}
\newcommand{\be}{\begin{equation}}
\newcommand{\ee}{\end{equation}}

\newcommand\lsim{\mathrel{\rlap{\lower4pt\hbox{\hskip1pt$\sim$}}
        \raise1pt\hbox{$<$}}}
\newcommand\gsim{\mathrel{\rlap{\lower4pt\hbox{\hskip1pt$\sim$}}
        \raise1pt\hbox{$>$}}}

\newcommand{\jcap}{{J.~Cosm.~Astrop.~Phys.}}

\newcommand{\aap}{{Astron.~Astrophys.}}
\newcommand{\apjl}{{Astrophys.~J.~Lett.}}
\newcommand{\apjs}{{Astrophys.~J.~Supp.}}
\newcommand{\aj}{{Astron.~J.}}
\newcommand{\mnras}{{Mon.~Not.~R.~Astron.~Soc.}}
\usepackage{graphicx}
\usepackage[large]{subfigure}
\usepackage{amssymb, amsmath}
\usepackage[amssymb]{SIunits}
\usepackage{epstopdf}
\usepackage{hyperref}
\usepackage{url}
\usepackage{aas_macros}
\usepackage{natbib}

\begin{document}
\title{Can Early Dark Energy Explain EDGES?}
\author{J.~Colin Hill\footnote{jch@ias.edu}}
\affiliation{School of Natural Sciences, Institute for Advanced Study, Princeton, NJ, USA 08540}
\affiliation{Center for Computational Astrophysics, Flatiron Institute, New York, NY, USA 10003}
\author{Eric J.~Baxter\footnote{ebax@sas.upenn.edu}}
\affiliation{Dept.~of Physics and Astronomy, University of Pennsylvania, Philadelphia, PA, USA 19104}
\begin{abstract}
The Experiment to Detect the Global Epoch of Reionization Signature (EDGES) collaboration has reported the detection of an absorption feature in the sky-averaged spectrum at $\approx 78$~MHz.  This signal has been interpreted as the absorption of cosmic microwave background (CMB) photons at redshifts $15 \lesssim z \lesssim 20$ by the 21cm hyperfine transition of neutral hydrogen, whose temperature is expected to be coupled to the gas temperature by the Wouthuysen-Field effect during this epoch.  Because the gas is colder than the CMB, the 21cm signal is seen in absorption.  However, the absorption depth reported by EDGES is more than twice the maximal value expected in the standard cosmological model, at $\approx 3.8\sigma$ significance.  Here, we propose an explanation for this depth based on ``early dark energy'' (EDE), a scenario in which an additional component with equation of state $w=-1$ contributes to the cosmological energy density at early times, before decaying rapidly at a critical redshift, $z_c$.  For $20 \lesssim z_c \lesssim 1000$, the accelerated expansion due to the EDE can produce an earlier decoupling of the gas temperature from the radiation temperature than that in the standard model, giving the gas additional time to cool adiabatically before the first luminous sources form.  We show that the EDE scenario can successfully explain the large amplitude of the EDGES signal.  However, such models are strongly ruled out by observations of the CMB temperature power spectrum.  Moreover, the EDE models needed to explain the EDGES signal exacerbate the current tension in low- and high-redshift measurements of the Hubble constant.  We conclude that non-finely-tuned modifications of the background cosmology are unlikely to explain the EDGES signal while remaining consistent with other cosmological observations.
\end{abstract}
\maketitle

\section{Introduction}
\label{sec:intro}
Recently, the Experiment to Detect the Global Epoch of Reionization Signature (EDGES) collaboration reported the detection of an absorption feature in the monopole sky spectrum centered at a frequency of $\approx 78$ MHz~\cite{EDGES}.  The signal was attributed to absorption of cosmic microwave background (CMB) photons by the 21cm spin-flip transition of neutral hydrogen at redshifts $z \approx 15$--20, during the epoch of the first star formation in the Universe.  During this epoch, Ly$\alpha$ photons from the first luminous sources couple the 21cm spin temperature to the gas temperature via the Wouthuysen-Field effect~\cite{Wouthuysen1952,Field1959}.  The gas is colder than the CMB at these redshifts, as it thermally decouples from the photon temperature at $z \approx 150$ and then cools adiabatically as $\propto (1+z)^2$, whereas the CMB temperature cools as $\propto (1+z)$. Consequently, the Wouthuysen-Field coupling causes the spin temperature to decrease, leading to a 21cm absorption feature during the era of the first star formation.  Eventually, heating due to the luminous sources increases the spin temperature to the point that the absorption vanishes, and the 21cm signal is seen in emission against the CMB at lower redshifts during reionization (see e.g.,~\cite{PL2008,PL2012} and references therein for additional details).

While the EDGES absorption signal appears to originate at roughly the
redshift range expected in many theoretical
models~(e.g.,~\cite{Cohen2017}), its amplitude and shape are rather
surprising.  In particular, the absorption amplitude ($\approx -500$
mK) is significantly (3.8$\sigma$) deeper than even the maximum value
possible in the thermal history of the standard $\Lambda$CDM cosmology
($\approx 200$ mK, if the spin temperature is fully coupled to the gas
temperature).  Taking the measurement at face value, two possible
explanations for the observed amplitude of the absorption feature are:
(i) the gas temperature, $T_{gas}$, at $z \approx 20$ was actually colder
than in the standard scenario, or (ii) the photon background
temperature, $T_{\gamma}$, was brighter than in the standard model at
$z \approx 20$.  In case (i), the observed signal implies $T_{gas}(z=20)
\approx 3.7$~K, roughly 2.5 times smaller than the minimum value
possible in the standard model, $T_{gas}(z=20) \approx 9.3$~K.  In case
(ii), the signal implies $T_{\gamma}(z=20) \approx 143$~K, roughly 2.5
times higher than in the standard scenario ($T_{\gamma}(z=20) \approx
57.2$~K), in which the CMB is the only relevant background radiation
during this epoch.

Both explanations have 
received attention in the literature.  Models based on dark matter --
baryon scattering have been proposed in order to cool the gas below
the minimum possible $\Lambda$CDM
temperature~\cite{Barkana2018Nature,ML2018}.  The dark matter is much
colder than the gas at these redshifts due to its much earlier
decoupling time from the thermal bath, and thus the scattering can
reduce $T_{gas}$.  However, the viable parameter space of these models is
severely constrained by other
observations~\cite{Berlin2018,Barkana2018,ML2018}.  Alternatively, models of
excess photon production at high redshift have been suggested, ranging
from purely phenomenological proposals~\cite{FH2018} (see also Ref.~\cite{Chluba2015} for a general treatment of the effects of photon injection at early times) to physical
models based on populations of obscured black
holes~\cite{Ewall-Wice2018}, soft photon emission from light dark
matter~\cite{Fraser2018}, or resonant oscillation of dark photons into regular photons in the Rayleigh-Jeans tail of the CMB~\cite{Pospelov2018}.

Here, we consider an interpretation of the EDGES signal in terms of an
unexpectedly low gas temperature at the redshift corresponding to the onset of the observed absorption feature.\footnote{We do not attempt to explain the shape of the absorption feature, but rather only the large depth at its onset.}  This effect
  can be achieved by pushing the time at which the gas and CMB
  temperatures decouple to earlier times than in the
  standard scenario (i.e., $z \gtrsim 150$).  In this case, the gas has sufficient time to cool adiabatically by $z
\approx 20$ to explain the amplitude of the EDGES signal.  In our model, the early
decoupling occurs as a result of ``early dark energy'' (EDE).  We model EDE as a
contribution to the energy density 
that has equation of state parameter $w = -1$ at high redshift, but then decays into a component with $w = 1$ (i.e., the equation of state for a free scalar field) rapidly at some critical redshift, $z_c$, in order to satisfy the range of existing cosmological constraints.  The EDE leads to an increase in the Hubble parameter
$H(z)$ at early times compared to its evolution in $\Lambda$CDM, which leads to the Compton-heating process
between gas and photons falling out of equilibrium at a higher
redshift than it does in $\Lambda$CDM.

EDE scenarios have attracted interest over the past decade, from both phenomenological and fundamental-physics perspectives (e.g.,~\cite{Doran-Robbers2006,Calabrese2011,Pettorino2013,KK2016,Shi-Baugh2016}).  Phenomenological interest has arisen due to hints of excess relativistic energy density in the pre-recombination Universe (which have largely vanished in most recent CMB data analyses)~\cite{Calabrese2011} and the possibility of explaining the current tension in inferences of the Hubble constant from CMB data and low-redshift astronomical data~\cite{KK2016}.  From the fundamental physics standpoint, string axiverse models predict the existence of many light (pseudo)scalar fields~\cite{Arvanitaki2010}, some of which could give rise to periods of accelerated expansion in earlier epochs of the Universe~\cite{Marsh2011,Kamionkowski2014}.  It is also possible that our vacuum decayed to its current (small) value from a higher-energy metastable state, e.g., in the context of the string landscape~(e.g.,~\cite{Bousso-Polchinski2000}).  In this paper, we adopt a purely phenomenological approach, and assess whether an EDE scenario could potentially explain the EDGES signal, while remaining consistent with other cosmological constraints.

Unless stated otherwise, we assume cosmological parameter values from the Planck 2015 ``TT+lowP+lensing'' analysis, e.g., $H_0 = 67.74$~km/s/Mpc, $\Omega_m = 0.3075$, and $\Omega_b = 0.0486$~\cite{Planck2015params}.

\section{Thermal History in Early Dark Energy Models}
\label{sec:EDE}

We adopt the phenomenological EDE model of Ref.~\cite{KK2016}, which in turn was inspired by string axiverse scenarios~\cite{Kamionkowski2014}.  Such scenarios rely on generic string theoretic predictions of the existence of $\mathcal{O}(100)$ axion-like fields, whose masses can span many orders of magnitude~\cite{Arvanitaki2010}.  At various points in cosmic history, each axion field can become dynamical and thereby drive a period of accelerated expansion; this depends on the initial value of each axion misalignment angle.  The picture is appealing in the sense that our current period of cosmic acceleration becomes only a one-in-100 chance occurrence, rather than a one-in-$10^{120}$ occurrence, as postulated in anthropic explanations involving the string landscape~\cite{Kamionkowski2014}.  While constraints on the model have been derived~(e.g.,~\cite{Poulin2018}), here we take an agnostic view about the fundamental physics responsible for this phenomenology.  For our purposes, we are interested only in the possibility of accelerated expansion in the epoch relevant to decoupling ($20 \lesssim z \lesssim 1000$).  In this model, the EDE contribution to the cosmological energy density is given by:
\be
\label{eq:EDE}
\rho_{ee}(a) = \rho_c \Omega_{ee} \left( \frac{1+a_c^6}{a^6+a_c^6} \right) \,,
\ee
where $a$ is the scale factor of the Universe, $\rho_c$ is the critical density at $z=0$, $\Omega_{ee} \equiv \rho_{ee}(a=1)/\rho_c$, and $a_c \equiv 1/(1+z_c)$ corresponds to the critical redshift at which the EDE transitions from $w=-1$ to $w = 1$ behavior.  The pressure of the EDE is given by:
\be
\label{eq:EDEpressure}
p_{ee}(a) = \rho_{ee}(a) \left( \frac{a^6-a_c^6}{a^6+a_c^6} \right) \,.
\ee
The EDE thus behaves as a cosmological constant ($w=-1$) at $z \gg z_c$ and as a free scalar field ($w=1$) at $z \ll z_c$.  We treat $z_c$ and $\Omega_{ee}$ as free parameters in the following.  For simplicity, our calculations only include changes to the background cosmology due to EDE, leaving out a full treatment of cosmological perturbation theory in these models (e.g.,~\cite{Caldwell1998}).  This approximation suffices for the observables considered below.\footnote{The effect of perturbations in the EDE fluid on the CMB temperature power spectrum is, in fact, non-negligible~\cite{Weller-Lewis2003}.  We comment on the implications for CMB-derived constraints in Sec.~\ref{sec:constraints}.}

We compute the thermal history of the Universe using the most recent version of Recfast~\cite{Seager1999,CT2011,Rubino-Martin2010,Chluba2010,CVD2010}.  We modify the code to implement the effects of EDE on the expansion history, which then alters the thermal history via the coupled differential equations describing the temperature and ionization evolution of the various components in the Universe.  Concretely, the gas temperature evolution is given
by~(e.g.,~\cite{Seager1999,MS2009}): \be
\label{eq:dTgdz}
\frac{dT_{gas}}{dz} = \frac{T_{gas}(z) - T_{\gamma}(z)}{(1+z) H(z) t_C(z)} + \frac{2 T_{gas}(z)}{(1+z)} \,,
\ee
where we assume the photon temperature $T_{\gamma} (z) \equiv T_{\rm CMB} (z) = 2.726 (1+z) \, {\rm K}$ and the Compton-heating timescale is
\be
\label{eq:tC}
t_C(z) = \frac{3 m_e c}{8 \sigma_T a_R T_{\gamma}^4(z)} \left(\frac{1 + f_{\rm He}(z) + x_e(z)}{x_e(z)} \right) \,.
\ee
Here, $m_e$ is the electron mass, $c$ is the speed of light, $\sigma_T$ is the Thomson scattering cross-section, $a_R$ is the radiation constant, $f_{\rm He}(z)$ is the fractional abundance of helium by number, and $x_e(z)$ is the free electron fraction normalized to the hydrogen number density.  The gas temperature decouples from the radiation temperature when $H \approx 1/t_C$, after which it cools adiabatically until the first luminous sources form.

\begin{figure}
\includegraphics[width=0.5\textwidth]{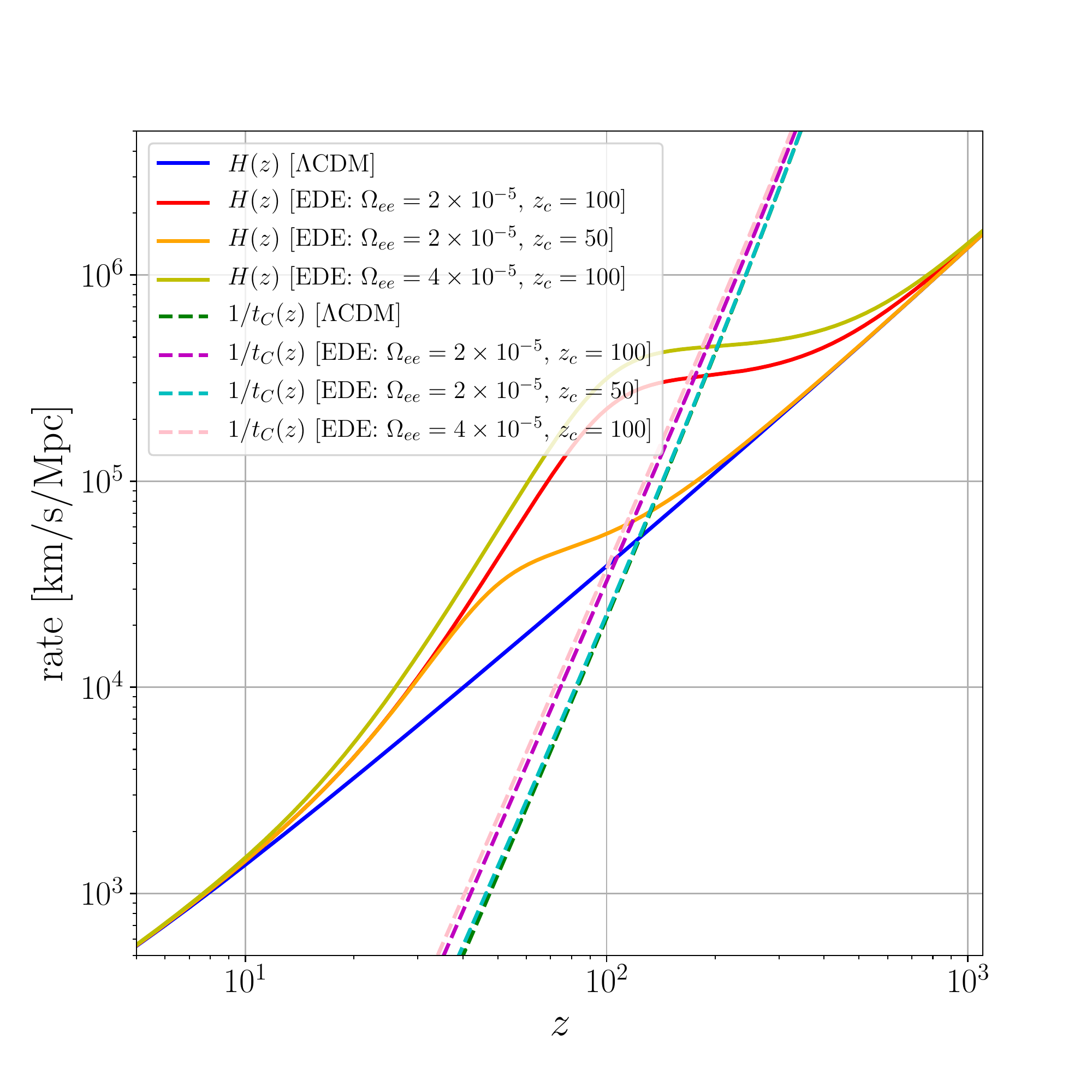}
\caption{\label{fig:rates} Expansion rate (solid curves) and Compton-heating rate (dashed curves) for standard $\Lambda$CDM (blue solid/green dashed) and various EDE models.  The red solid/magenta dashed curves show the rates for $\Omega_{ee}=2\times 10^{-5}$ and $z_c=100$; the orange solid/cyan dashed curves show the rates for $\Omega_{ee}=2\times 10^{-5}$ and $z_c=50$; and the yellow solid/pink dashed curves show the rates for $\Omega_{ee}=4\times 10^{-5}$ and $z_c=100$.  The decoupling of the gas temperature from the radiation temperature occurs when $H(z) \approx 1/t_C(z)$, i.e., when Compton-heating falls out of equilibrium.  In the EDE models, decoupling occurs earlier, and thus the gas temperature is lower at late times than it is in $\Lambda$CDM.}
\end{figure}

\begin{figure}
\includegraphics[width=0.5\textwidth]{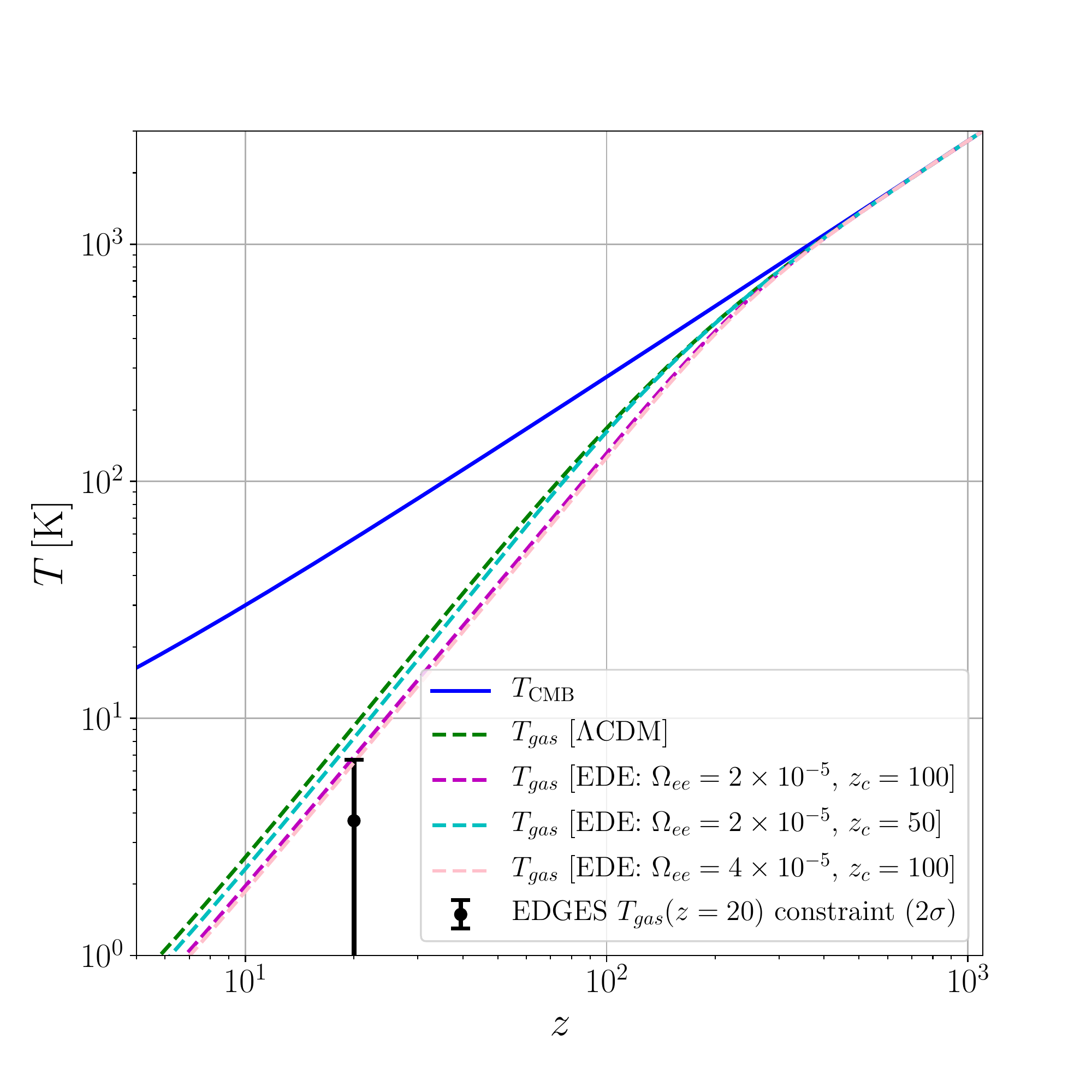}
\caption{\label{fig:temps} Gas temperature (dashed curves) and radiation temperature (solid blue curve) for standard $\Lambda$CDM and various EDE models.  The radiation temperature evolution is identical in all models, and thus we show it only in $\Lambda$CDM for clarity.  The dashed green curve shows the gas temperature evolution for $\Lambda$CDM; the dashed magenta curve shows the same quantity for an EDE model with $\Omega_{ee}=2\times 10^{-5}$ and $z_c=100$; the dashed cyan curve shows the same quantity for an EDE model with $\Omega_{ee}=2\times 10^{-5}$ and $z_c=50$; and the dashed pink curve shows the same quantity for an EDE model with $\Omega_{ee}=4\times 10^{-5}$ and $z_c=100$.  The black point indicates the gas temperature at $z\approx 20$ implied by the EDGES signal, assuming the standard CMB radiation background.  In the EDE models shown here, the gas temperature is lower at this redshift than in $\Lambda$CDM, due to the earlier decoupling of the gas from the radiation.}
\end{figure}

In order to explain the EDGES signal, the EDE must significantly impact $H(z)$ in the redshift window relevant to decoupling, i.e., $20 \lesssim z \lesssim 1000$.  In Fig.~\ref{fig:rates}, we show $H(z)$ and the Compton-heating rate $1/t_C(z)$ in $\Lambda$CDM and in three EDE models with various $\Omega_{ee}$ and $z_c$ values, which have been selected for their significant impact on the thermal evolution of the gas.  The EDE leads to a period in which $H(z)$ is approximately constant, thus yielding an earlier epoch at which $H(z) \approx 1/t_C(z)$ compared to that in $\Lambda$CDM.  As a result, the gas thermally decouples from the radiation earlier than it would in the standard scenario.  The gas then adiabatically cools, and is colder at $z \approx 20$ than it would be in the standard model.  It can also be seen in Fig.~\ref{fig:rates} that there is a small change to the evolution of $1/t_C(z)$ itself, but the impact of this change on the decoupling time is subdominant to that of the change in $H(z)$.

In Fig.~\ref{fig:temps}, we show the modified gas temperature evolution in the EDE models compared to that in $\Lambda$CDM.  As expected, the earlier decoupling in these models leads to a lower gas temperature at $z \approx 20$ than in $\Lambda$CDM.  For the models shown in Fig.~\ref{fig:temps}, we find $T_{gas}(z=20) = 6.9$~K ($\Omega_{ee}=2\times 10^{-5}$ and $z_c=100$); $T_{gas}(z=20) = 8.3$~K ($\Omega_{ee}=2\times 10^{-5}$ and $z_c=50$); and $T_{gas}(z=20) = 6.5$~K ($\Omega_{ee}=4\times 10^{-5}$ and $z_c=100$).  These models thus lie within $\lesssim 2$--$3\sigma$ of the gas temperature at $z \approx 20$ implied by the EDGES measurement.  In general, for values of $z_c$ in the range $50 \lesssim z_c \lesssim 400$, increasing $\Omega_{ee}$ decreases $T_{gas}(z=20)$, bringing the predicted value into further closer agreement with EDGES.  We conclude from these calculations that the EDE scenario is capable of explaining the amplitude of the EDGES absorption signal at $z=20$.

\section{Constraints from Other Cosmological Observables}
\label{sec:constraints}

We now consider constraints on the EDE scenario from other observables, primarily those related to the CMB.  The EDE modifies the expansion history of the universe.  By construction, we avoid any violations of low-redshift ($z \lesssim 4$) constraints on the distance-redshift relation (e.g., from supernovae or baryon acoustic oscillations) by restricting $z_c \gtrsim 10$.  The only constraint on the expansion history at higher redshifts is that due to the precise measurement of the acoustic scale by CMB experiments, which is directly related to the angular diameter distance to the surface of last scattering, $D_{SLS}$.\footnote{Constraints on the expansion rate at much higher redshifts can also be obtained from Big Bang Nucleosynthesis (BBN).  For the EDE models considered here to explain the EDGES signal, the Universe is fully radiation-dominated during BBN, so no relevant EDE constraints are expected from this observable.}   The CMB thus provides an integral constraint on $H^{-1}(z)$ to $z \approx 1100$.  For $z_c < 1100$, the EDE model decreases $D_{SLS}$, since $H(z)$ is increased relative to its $\Lambda$CDM values in the regime where the EDE non-negligibly contributes to the cosmic energy density.

\begin{figure}
  \includegraphics[width=0.5\textwidth]{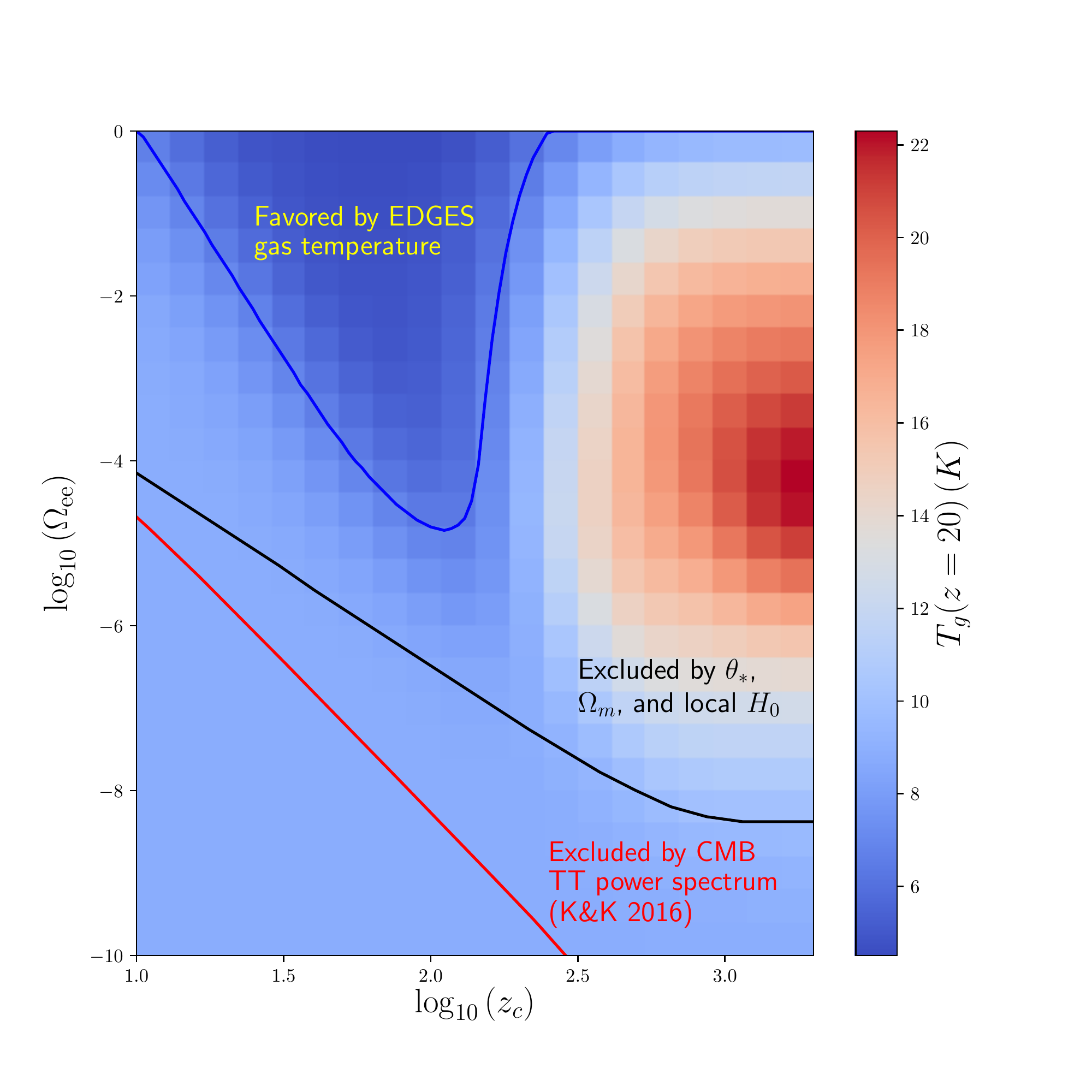}
  \caption{\label{fig:exclusion} Constraints in the EDE $(\Omega_{ee}, z_{c})$ parameter space from cosmological observables compared to the region necessary to explain the EDGES signal.  The region above the red curve is excluded by Planck measurements of the CMB TT power spectrum~\cite{KK2016} ($1\sigma$ exclusion), primarily because of the large ISW contribution associated with the EDE epoch.  The region above the black curve is excluded by purely geometric constraints from the CMB (i.e., the angular size of the acoustic scale) when combined with local measurements of $H_0$ and the \citet{DESy1} constraint on $\Omega_m$.  Given concerns about tension in measurements of $H_0$, we adopt a conservative $4\sigma$ exclusion limit for the black curve.  The region above the blue curve contains the models necessary to explain the EDGES signal at $2\sigma$ by lowering the gas temperature at $z \approx 20$ compared to that in $\Lambda$CDM.  The color bar indicates the gas temperature at $z=20$ for each point in parameter space.  The necessary models are strongly excluded (note the logarithmic scales on the axes).}
\end{figure}

In order to satisfy the CMB constraint on $D_{SLS}$ in the EDE scenario, we must either decrease $H_0$ or ``compensate'' the increased $H(z)$ in the EDE-dominated epoch with a decreased $H(z)$ in some other era of cosmic history.  In the context of flat $\Lambda$CDM + EDE models, the only way to make such a change to $H(z)$ while preserving $H_0$ is to change $\Omega_m$.  However, $\Omega_m$ is effectively fixed by low-redshift measurements, such as the \citet{DESy1} constraint: $\Omega_m = 0.264^{+0.032}_{-0.019}$.  Given the tightness of these constraints, we can ignore uncertainty in $\Omega_m$ and simply fix $\Omega_m = 0.264$.  Then, to compensate for any change in $H(z)$ due to EDE, $H_0$ must be varied.  
For the models shown in Figs.~\ref{fig:rates}~and~\ref{fig:temps}, the necessary value of $H_0$ in order to preserve $D_{SLS}$ is: $H_0 = 59.26$~km/s/Mpc for $\Omega_{ee}=2\times 10^{-5}$ and $z_c=100$; $H_0 = 62.57$~km/s/Mpc for $\Omega_{ee}=2\times 10^{-5}$ and $z_c=50$; and $H_0 = 57.25$~km/s/Mpc for $\Omega_{ee}=4\times 10^{-5}$ and $z_c=100$.  Thus, $H_0$ must be decreased significantly from the best-fit value found in $\Lambda$CDM fits to the Planck data.  In the absence of additional changes to the background cosmology (i.e., beyond flat $\Lambda$CDM + EDE) that could increase the inferred $H_0$ value from CMB data, the EDE models necessary to explain the EDGES signal therefore significantly exacerbate the current tension between low- and high-redshift constraints on $H_0$~\cite{Planck2015params,Riess2018}.\footnote{Note that for EDE models with $z_c > 1100$, the $H_0$ tension can be moderately \textit{reduced}~\cite{KK2016}, but these models cannot explain the EDGES signal.}

Fig.~\ref{fig:exclusion} shows constraints on the EDE parameters derived from these arguments (the area above the black curve), as well as the region in EDE parameter space that would be necessary to satisfy the observed amplitude of the EDGES absorption feature within $2\sigma$ (the area above the blue curve).  The region above the black curve is excluded by the requirement that $H_0 > 66.84$~km/s/Mpc, i.e., that it is consistent with the constraint from Ref.~\cite{Riess2018} within $4\sigma$, and that $D_{SLS}$ is unchanged from its value in our fiducial cosmology.   We adopt a conservative $4\sigma$ limit here in order to obviate any concerns due to the current tension in low- and high-redshift constraints on $H_0$.  We fix $\Omega_m = 0.264$ (i.e., the best-fit value from Ref.~\cite{DESy1}) since uncertainty on $\Omega_m$ is subdominant to that on $H_0$.  Also, our approach ignores uncertainty in the current measurement of the acoustic scale, as the effect of this uncertainty is again subdominant compared to that of the uncertainty in $H_0$.  From Fig.~\ref{fig:exclusion}, it is clear that these geometric probes alone suffice to rule out the EDE parameter space required to explain the EDGES signal.

However, the CMB offers additional constraining power on the EDE scenario, beyond the $D_{SLS}$ constraint described above.  In particular, the EDE contributes to the CMB temperature (TT) power spectrum via the integrated Sachs-Wolfe (ISW) effect and slightly modifies the recombination history as well.  The small-scale TT power spectrum is also sensitive to $H(z)$ around the epoch of recombination, which further constrains the EDE scenario.  The quantitative details of these effects (and the change to $D_{SLS}$) depend on the EDE model considered, particularly the value of $z_c$.  This scenario was considered in Ref.~\cite{KK2016}, who performed an approximate Fisher analysis of the Planck TT power spectrum to constrain the model.  Their analysis was motivated by an attempt to resolve the tension between constraints on $H_0$ from direct astronomical measurements and those from other cosmological observables.  Relaxing the $H_0$ tension generally requires $z_c > 1100$ in the EDE model, in contrast to the post-recombination $z_c$ values needed to explain the EDGES signal.

We extract relevant constraints on $\Omega_{ee}$ and $z_c$ from Fig.~13 of Ref.~\cite{KK2016}, in which the authors give the $1\sigma$ upper limit on the fractional energy density contributed by the EDE component at its critical redshift $z_c$.  Although the optical depth $\tau$ is held fixed in their analysis, the other cosmological parameters are marginalized over (at the Fisher level), and the resulting constraints are so strong that marginalizing $\tau$ would not change our conclusions.  Fig.~\ref{fig:exclusion} shows the region of the $(\Omega_{ee}, z_{c})$ parameter space that is excluded by the TT power spectrum (the entire area above the red curve).  It is evident that the CMB constraints strongly rule out the EDE scenarios needed to cool the gas temperature sufficiently to explain the EDGES data.  This constraint arises from the significant production of early ISW power during the EDE epoch and the change to the angular size of the CMB acoustic scale.  Although the constraints from Ref.~\cite{KK2016} are approximate in nature, they are consistent with those from more precise analyses of other EDE models~\cite{Doran-Robbers2006,Calabrese2011,Pettorino2013}, and suffice to strongly rule out the EDE models needed to explain the EDGES signal.\footnote{We note that the analysis in Ref.~\cite{KK2016} neglects EDE perturbations, which is unphysical; inclusion of this effect substantially alters the predicted CMB TT power spectrum in the EDE model~(A.~Lewis, priv.~comm.).  In particular, a characteristic scale corresponding to the horizon size at the time of EDE domination is imprinted on the power spectrum, which changes the ISW signature associated with the EDE.  Correctly accounting for this effect modifies the constraints from Ref.~\cite{KK2016}.  The allowed region in Fig.~\ref{fig:exclusion} actually {\it decreases} with this change~(A.~Lewis, priv.~comm.).  
The conclusions of Ref.~\cite{KK2016} regarding $H_0$ in this EDE scenario are also modified, with substantially less freedom in the allowed change to $H_0$~(A.~Lewis, priv.~comm.).  Consequently, the qualitative conclusions for the EDGES interpretation implied by Fig.~\ref{fig:exclusion} are effectively unchanged after accounting for the effects of EDE perturbations: the values of the EDE parameters required to explain the EDGES signal remain ruled out by multiple orders of magnitude.}

Finally, we note that the EDE scenarios considered here also suppress the growth of structure, since the Universe is no longer fully matter-dominated from the epoch of matter-radiation equality to that of matter-$\Lambda$ equality.  This effect could actually pose a problem for explaining the EDGES signal, since it is necessary for the first luminous sources to have formed by $z \approx 20$ to obtain the measured 21cm absorption signal.  Given the strong constraints already found above, we do not investigate this issue quantitatively here.

\section{Conclusions}
\label{sec:conclusions}
It is highly unlikely that non-finely-tuned modifications of the background cosmology, such as the simple EDE models considered here, can explain the EDGES observation.  Strong constraints arise from ISW contributions to the CMB TT power spectrum, changes to the angular size of the CMB acoustic scale, and effects on the Hubble rate around the time of recombination.  Moreover, the necessity of decreasing $H_0$ to preserve $D_{SLS}$ in these models further exacerbates the tension in this parameter in current cosmological data analyses.  Additional broadening of the cosmological model to include curvature, additional relativistic species, or other contributions could potentially alleviate some of these constraints (e.g., $H_0$), but such proposals are unlikely to succeed.  For a concrete example, if we consider the EDE model with $\Omega_{ee}=4\times 10^{-5}$ and $z_c=100$ (which yields $H_0 = 57.25$~km/s/Mpc as discussed in Sec.~\ref{sec:constraints}), then we can introduce additional relativistic degrees of freedom in the early Universe ($\Delta N_{\rm eff}$) to bring $H_0$ back to a value that is realistically plausible in light of direct measurements (i.e., $H_0 \approx 70$~km/s/Mpc).  An approximate estimate (e.g., following the reasoning in Ref.~\cite{Riess2016}) indicates that we would require $\Delta N_{\rm eff} \approx 2$, which is in strong disagreement with constraints from the damping tail of the CMB power spectrum (albeit derived in the context of the minimal $\Lambda$CDM + $N_{\rm eff}$ model)~\cite{Planck2015params}.  Thus, it is highly unlikely that EDE, even in combination with further broadening of the cosmological model, can satisfy all relevant observational constraints while also explaining the EDGES signal.  As seen in Fig.~\ref{fig:exclusion}, such proposals are extremely difficult to reconcile with the precisely measured CMB TT power spectrum.

\begin{acknowledgments}
We thank Adam Lidz, David Spergel, and Matias Zaldarriaga for useful conversations.  We are grateful to Jens Chluba for providing information about the Recfast code.  We also thank Antony Lewis for his independent work in assessing the robustness of the CMB constraints.  JCH acknowledges support from the Friends of the Institute for Advanced Study. EB is partially supported by the US Department of Energy grant DE-SC0007901 and funds from the University of Pennsylvania.
\end{acknowledgments}

\pagebreak


\end{document}